\documentclass[aps,showpacs,prb,reprint,longbibliography,superscriptaddress]{revtex4-2}

\usepackage[colorlinks=true,linkcolor=blue,citecolor=blue,urlcolor=blue]{hyperref}
\usepackage{setspace} 
\usepackage{graphicx}
\usepackage{amsmath}
\usepackage{color}
\usepackage{amsmath}
\usepackage{amssymb}
\usepackage{verbatim}
\usepackage{physics}
\usepackage{latexsym}
\usepackage{enumerate} 
\usepackage{bm} 
\usepackage[caption=false]{subfig}
\usepackage{multirow}
\usepackage{booktabs}
\usepackage{siunitx}

\usepackage{float}

\usepackage{lipsum}

\begin{document}

\title{\textbf{\Large{Phonon-induced localization of excitons in molecular crystals \\from first principles}}}

\author{Antonios M. Alvertis}
\email{amalvertis@lbl.gov}
\affiliation{Materials Sciences Division, Lawrence Berkeley National Laboratory, Berkeley, California 94720, USA}
\affiliation{Department of Physics, University of California Berkeley, Berkeley, United States}
\author{Jonah B. Haber}
\affiliation{Department of Physics, University of California Berkeley, Berkeley, United States}
\author{Edgar A. Engel}
\affiliation{Cavendish Laboratory, University of Cambridge, J.\,J.\,Thomson Avenue, Cambridge CB3 0HE, United Kingdom}
\author{Sahar Sharifzadeh}
\affiliation{Division of Materials Science and Engineering, Boston University, United States}
\affiliation{Department of Electrical and Computer Engineering, Boston University, United States}
\author{Jeffrey B. Neaton}
\email{jbneaton@lbl.gov}
\affiliation{Materials Sciences Division, Lawrence Berkeley National Laboratory, Berkeley, California 94720, USA}
\affiliation{Department of Physics, University of California Berkeley, Berkeley, United States}
\affiliation{Kavli Energy NanoScience Institute at Berkeley, Berkeley, United States}

\date{\today}
\begin{abstract}
The spatial extent of excitons in molecular systems underpins their photophysics and utility for optoelectronic applications.
Phonons
are reported to lead to both exciton localization and delocalization. However, a microscopic understanding of phonon-induced (de)localization is lacking, in particular how localized states form, the role of
specific vibrations, and
the relative importance of quantum and thermal nuclear fluctuations. Here we present a
first-principles study of these phenomena in solid pentacene, a prototypical molecular crystal, capturing the formation of bound excitons, exciton-phonon coupling to all orders, and phonon anharmonicity, using density functional theory, the \emph{ab initio} $GW$-Bethe-Salpeter equation approach, finite difference, and path integral techniques. We find that for pentacene zero-point nuclear motion causes uniformly strong localization, with thermal motion providing additional localization only for Wannier-Mott-like excitons. Anharmonic effects drive temperature-dependent localization, and while such effects prevent the emergence of highly delocalized excitons, we explore the conditions
under which these might be realized.

\end{abstract}

\maketitle

\emph{Introduction.}-- 
Photoexcitation of organic molecular crystals leads to strongly bound electron-hole pairs, or excitons, due to the weak screening of the Coulomb interaction in these systems. Depending on factors
such as the size of the molecular building blocks and the spin of the electron-hole pair, exciton radii can vary from those of localized Frenkel excitons~\cite{PhysRev.37.17,PhysRev.37.1276} to spatially
extended excitons that approach the Wannier-Mott limit~\cite{PhysRev.52.191,TF9383400500,Cudazzo2012,Cudazzo2013,Cudazzo2015}. The spatial
extent of these excited states is important to 
applications of organic semiconductors such as photovoltaics~\cite{Distler2021} and LEDs~\cite{Reineke2009}, since it affects properties including
the nature of their interaction with phonons~\cite{alve+20prb}, their transport~\cite{Arago2015} and non-radiative recombination~\cite{Kupgan2021}.

Critical to affecting the spatial extent of
excited states are lattice vibrations, which are
generally thought to result in wavefunction localization~\cite{PhysRev.109.1492}. Phonons  can strongly renormalize one- and two-particle excitation energies of organic systems, influencing the optical gap and the charge carrier mobility~\cite{Brown-Altvater2020,Schweicher2019,alve+20prb}. Phonons in these systems have generally been thought to 
lead to localized excitons that diffuse via, \emph{e.g.},
a F\"{o}rster or Dexter mechanism~\cite{Athanasopoulos2009,SudhaDevi2008}. However, it has recently
been proposed that in certain well-ordered organic crystals atomic motion can give rise to
configurations that favor strong transient exciton delocalization, having a beneficial effect
to transport~\cite{Sneyd2021,Giannini2022,Sneyd2022}. This
transient exciton delocalization is similar to transient \emph{charge} delocalization~\cite{Fratini2016,Zhang2014,doi:10.1146/annurev-physchem-042018-052353}, wherein phonons lead to configurations with large overlaps
between neighboring molecular orbitals~\cite{Troisi2005}
and hence highly delocalized states~\cite{Giannini2019}.


Despite these insights, a rigorous microscopic understanding of phonon-induced modulations to exciton radii, one that accounts for electron-hole interactions,
strong exciton-phonon coupling at finite temperatures~\cite{Monserrat2015,alve+20prb}, and the anharmonicity of low-frequency motions in
molecular crystals~\cite{Alvertis2022,Seiler2020,Fetherolf2022,Rossi2016}, is still lacking.
Here we elucidate the microscopic mechanism of exciton localization in extended molecular solids. We employ a first-principles computational framework which captures all aforementioned effects, combining density functional theory (DFT), the Green's function-based \emph{ab initio} $GW$-Bethe Salpeter equation (BSE) approach for accurately describing exciton effects~\cite{Rohlfing2000}, finite-difference methods for
strong exciton-phonon interactions~\cite{Monserrat2018,alve+20prb}, and path 
integral techniques for describing phonon anharmonicity~\cite{Kapil2016,ceri+10jcp}. We apply this framework to the prototypical molecular crystal pentacene and 
show that zero-point nuclear motion leads to strong localization of singlet and triplet excitons, reducing their average electron-hole separation by more than a factor of two.
Temperature increases further reduce the size of delocalized Wannier-Mott-like excitons, an effect driven by anharmonic phonons. The trends in exciton radii
are reflected in the dispersion of their energies in reciprocal space.  While highly
delocalized excitons do appear at large phonon displacements, anharmonicity reduces the amplitude associated with these motions, suppressing
transient delocalization for exciton transport.

\begin{figure}[tb]
    \centering
    \includegraphics[width=0.9\linewidth]{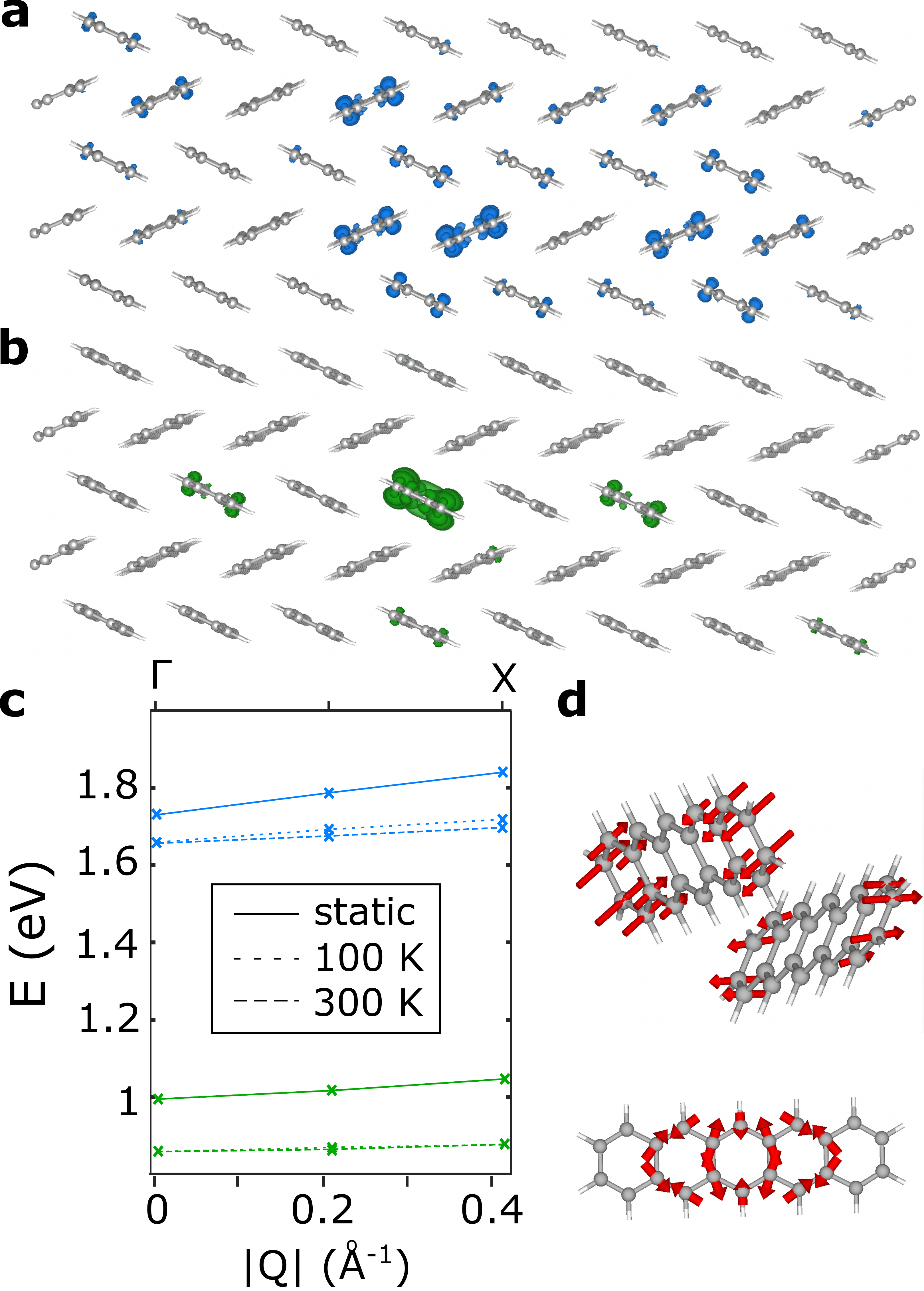}
    \caption{Isosurfaces of electron distributions of singlet (blue, panel \textbf{a}) and triplet (green, panel \textbf{b}) excitons for a hole fixed at the center of the plotted area, and corresponding dispersions (panel \textbf{c}, same color scheme) in molecular crystals. A typical low-frequency (top) and high-frequency (bottom) phonon of pentacene is shown in panel \textbf{d}.}
    \label{fig:dispersions}
\end{figure}

\emph{System and methods.}-- 
We focus on the widely studied molecular crystal pentacene~\cite{Haas2007}, which hosts a delocalized Wannier-Mott-like singlet exciton (Fig.\,\ref{fig:dispersions}a)
and a more localized Frenkel-like triplet exciton (Fig.\,\ref{fig:dispersions}b)~\cite{Refaely-Abramson2017,alve+20prb,Cudazzo2015}, for which
the effect of phonons is expected to be different. We compute excitons with principal quantum number $S$ and center-of-mass momentum $\boldsymbol{Q}$ using \emph{ab initio} DFT and $GW$-BSE
calculations with the Quantum Espresso~\cite{QE} and BerkeleyGW~\cite{Deslippe2012} codes. This involves
constructing the electron-hole kernel $K^{e-h}$ and solving the BSE~\cite{Rohlfing2000,Qiu2021} in reciprocal space in the electron-hole basis, namely

\begin{align}
    \label{eq:BSE}
    (E_{c\boldsymbol{k}+\boldsymbol{Q}}-E_{v\boldsymbol{k}})A^S_{cv\boldsymbol{k}\boldsymbol{Q}}\\ \nonumber +\sum_{c'v'\boldsymbol{k}'}\bra{c\boldsymbol{k}+\boldsymbol{Q},v\boldsymbol{k}}K^{e-h}\ket{c'\boldsymbol{k}'+\boldsymbol{Q},v'\boldsymbol{k}'}A^S_{c'v'\boldsymbol{k}'\boldsymbol{Q}}\\ \nonumber =\Omega^S_{\boldsymbol{Q}}A^S_{cv\boldsymbol{k}\boldsymbol{Q}},
\end{align}

\noindent
with input from prior DFT and $GW$ calculations. 
In Eq.~\ref{eq:BSE} the indices $c,v$ define conduction and valence states respectively, $\boldsymbol{k}$ is the crystal momentum, and $A^S_{cv\boldsymbol{k}\boldsymbol{Q}}$ is the amplitude contributed
by states $c,v$ with momentum $\boldsymbol{k}$ to the exciton with
momentum $\boldsymbol{Q}$. The exciton wavefunction can be written as

\begin{equation}
    \label{eq:exciton}
    \Psi_S^{\boldsymbol{Q}}(\boldsymbol{r}_e,\boldsymbol{r}_h)=\sum_{cv\boldsymbol{k}}A^S_{cv\boldsymbol{k}\boldsymbol{Q}}\psi_{c\boldsymbol{k}+\boldsymbol{Q}}(\boldsymbol{r}_e)\psi^*_{v\boldsymbol{k}}(\boldsymbol{r}_h),
\end{equation}


\noindent
where $\psi_{n\boldsymbol{k}}$ are the Kohn-Sham wavefunctions. The kernel $K^{e-h}$
consists only of an attractive `direct' term between electrons and holes for triplets,
while for singlets it also includes a repulsive `exchange' term,
giving singlets their greater spatial extent~\cite{Cudazzo2015,Rohlfing2000}.
The energies of the conduction and valence bands in Eq.\,\ref{eq:BSE} are obtained
within the so-called $GW$ approximation~\cite{Hybertsen1986} from self-energy corrections to DFT Kohn-Sham eigenvalues. This
approach has been shown to give highly accurate
descriptions of excitons in molecular crystals~\cite{Refaely-Abramson2017,alve+20prb,Rangel2016,Cudazzo2015,Lettmann2021}. The computational details for our DFT and $GW$-BSE calculations are
given in Supplemental Material~\cite{supp} Section\,S1. 

We treat the effect of phonons following Monserrat~\cite{Monserrat2018,Monserrat2016,Monserrat2016_2}, and in a manner similar in spirit to
Zacharias and Giustino~\cite{Zacharias2016,Zacharias2020}. For an
observable  $\mathcal{O}$ at a temperature $T$, we compute the
ensemble-average in the adiabatic approximation as

\begin{equation}
    \left\langle \mathcal{O}(T) \right\rangle_{\cal H}
    = \frac{1}{Z} \int dX \mathcal{O}(X) e^{ -\beta \cal H },
    \label{eq:ens_avg}
\end{equation} 

\noindent
where the canonical partition function $Z = \int dX e^{ -\beta \cal H }$ involves the configuration space integral $\int dX$~\cite{Patrick2014}. Non-adiabatic effects to the electron-phonon interactions of organic systems such as pentacene are
negligible~\cite{Miglio2020}.

The Hamiltonian $\cal H$ of the system includes electronic and nuclear degrees of freedom in general, and may be approximated
at different levels. One approach is to assume nuclear motion to be harmonic, reducing the phonon contribution to the Hamiltonian to the following form,

\begin{equation}
    {\cal H}^{\textrm{har}} 
    \equiv 
    \frac{1}{2} \sum_{n,{\bf q}}( \nabla_{u_{n,{\bf q}}}^2 +
    \omega_{n,{\bf q}}^2 u_{n,{\bf q}}^2),
\label{eq:har_H}
\end{equation}
in atomic units. Here, phonons of frequencies $\omega$ are labeled by their branch index $n$ and wavevector ${\bf q}$. We compute the ensemble-average $\left\langle \mathcal{O}^{\textrm{har}} \right\rangle$ in the Born-Oppenheimer approximation, tracing out all electronic degrees of freedom, using a finite-displacements approach~\cite{Kresse1995, Parlinski1997} to calculate phonon frequencies $\{ \omega_{n,{\bf q}} \}$ and eigendisplacements $\{ u_{n,{\bf q}} \}$, and then drawing $N$ random samples $\{ X^{\textrm{har}}_i \}$ from the multivariate Gaussian phonon distribution and calculating
the observables of interest $\{ \mathcal{O}(X^{\textrm{har}}_i) \}$. 
$\left\langle \mathcal{O}^{\textrm{har}} \right\rangle$ is then simply computed as the average of its value at the samples

\begin{equation}
    \left\langle \mathcal{O}^{\textrm{har}} \right\rangle
    = 
    \lim_{N \rightarrow \infty} \frac{1}{N}\sum_{i=1}^{N} \mathcal{O}(X^{\textrm{har}}_i).
    \label{eq:har_band_gap}
\end{equation}

\noindent
Eqs.\,\ref{eq:har_H} and\,\ref{eq:har_band_gap} are exact apart from the adiabatic and harmonic approximations, and the description
of phonon effects on any observable $\mathcal{O}$ in Eq.\,\ref{eq:har_band_gap} is non-perturbative~\cite{Monserrat2015}.

The use of the harmonic approximation in molecular crystals can lead
to unphysical results, due to highly anharmonic behavior of low-frequency phonons~\cite{Alvertis2022,Fetherolf2022}. In this work, we account for this anharmonicity by employing path-integral molecular dynamics (PIMD) which are rendered computationally tractable using the surrogate machine-learning (ML) potential $V^{\textrm{ML}}$ from Refs.~\cite{Kapil2022,Alvertis2022}, constructed to reproduce the potential energy surface (PES) from first-principles density functional theory (DFT) calculations. The modified phonon Hamiltonian

\begin{equation}
    {\cal H}^{\textrm{anhar}} \equiv \sum_{i=1}^{N_a} \frac{{\hat{\bf p}}_i^2}{2 m_i} + V^{\textrm{ML}}({\hat{\bf r}}_1,\ldots,{\hat{\bf r}}_{N_a})
\end{equation}
is used to run PIMD simulations at reduced computational cost, for a cell of $N_a$ atoms, with nucleus $i$ having a mass $m_i$, and $\hat{\bf p}_i$, $\hat{\bf r}_i$ its momentum and position operators respectively. We then draw random samples from the PIMD trajectories, and use these to compute vibrational averages of observables, analogously to Eq.\,\ref{eq:har_band_gap}, namely

\begin{equation}
    \left\langle \mathcal{O}^{\textrm{anhar}} \right\rangle
    = 
    \lim_{N \rightarrow \infty} \frac{1}{N}\sum_{i=1}^{N} \mathcal{O}(X^{\textrm{anhar}}_i).
    \label{eq:anhar_band_gap}
\end{equation}
Our simulations use a $2\times1\times1$ supercell of pentacene ($N_a=144$ atoms), capturing the effect of phonons at $\Gamma$ and at the band-edge $X$ on observables. Phonons beyond $\Gamma$ and $X$ have a minor effect on pentacene optical properties as discussed in Supplemental Material~\cite{supp} Section\,S1.C. 

To quantify exciton localization, we study two observables  $\mathcal{O}$. 
The first are the exciton 
energies at finite  center-of-mass momentum, $\Omega^S_{\boldsymbol{Q}}$, obtained through solving the BSE (Eq.\,\ref{eq:BSE}). The second is the average electron-hole separation for each excitation $S$, which we refer to as the exciton radius $r_{\text{exc}}$. This is obtained by post-processing
the BSE solution $\Psi_S$, as discussed elsewhere~\cite{Sharifzadeh2013} and in Supplemental Material~\cite{supp} Section\,S1. To determine the exciton radius, we compute the electron-hole correlation function as defined in Ref.~\cite{Sharifzadeh2013}, namely

\begin{equation}
    \label{eq:correlation_function}
    F_S(\boldsymbol{r})=\int_{V}d\boldsymbol{r}_h |\Psi^{\boldsymbol{Q}=0}_S(\boldsymbol{r}_e=\boldsymbol{r}_h+\boldsymbol{r},\boldsymbol{r}_h)|^2,
\end{equation}
where $V$ the volume of the primitive cell. $F_S(\boldsymbol{r})$ describes the probability of finding the electron-hole pair at a distance of $\boldsymbol{r}=\boldsymbol{r}_e-\boldsymbol{r}_h$, and
is computed as a discrete sum over hole positions. The average exciton radius for a given atomic configuration
is then

\begin{equation}
\label{eq:rexc}
r_{\text{exc}}=\int d|\boldsymbol{r}|F_S(|\boldsymbol{r}|)|\boldsymbol{r}|.
\end{equation}

\begin{table*}[tb]
\centering
  \setlength{\tabcolsep}{6pt} 
\begin{tabular}{cccccc}
\hline
 & $W^{\text{anhar}}(\text{S}_1)$ (meV) & $W^{\text{har}}(\text{S}_1)$ (meV) & $W^{\text{anhar}}(\text{T}_1)$ (meV) & $\Delta^{\text{anhar}}(\text{S}_1)$ (meV) & $\Delta^{\text{exp}}(\text{S}_1)$ (meV)~\cite{Graf2022}  \\
\hline
static & $110$ & $110$ & $52$ & $80$ & $---$ \\
$100$\,K & $59$ & $67$ & $18$ & $43$ & $44$ \\
$300$\,K & $41$ & $76$ & $19$ & $30$ & $23$ \\
\hline
\end{tabular}
\caption{The effect of phonons on the dispersion width $W=\Omega(X)-\Omega(\Gamma)$ for the first singlet $\Omega_S$ and triplet $\Omega_T$ excitons of pentacene, and on the width $\Delta=\Omega(\mathbf{Q}=0.4\hspace{0.1cm}\text{\AA}^{-1})-\Omega(\mathbf{Q}=0.1\hspace{0.1cm}\text{\AA}^{-1})$ for the singlet.}
\label{table:widths}
\end{table*}

Having described the main quantities in our computational framework, we may summarize it as follows. We generate displaced
configurations $X^{\textrm{har}}_i$ within the
harmonic approximation using a finite differences
approach, and $X^{\textrm{anhar}}_i$ within the anharmonic distribution through PIMD employing a previously-developed ML potential. The \emph{ab initio} BSE, Eq.\,\ref{eq:BSE}, is solved at these configurations,
followed by a calculation of the exciton radius via Eq.\,\ref{eq:rexc}. We then compute the vibrational
averages using Eqs.\,\ref{eq:har_band_gap} and\,\ref{eq:anhar_band_gap}.
Details of the convergence of the vibrational averages, the 
ML potential, and PIMD simulations, are given in Supplemental Material~\cite{supp} Section\,S1. 

\emph{Results.}-- We first discuss exciton properties obtained from solving the
BSE without consideration of phonons. We refer to these clamped-ion solutions as the `static' case. Fig.\,\ref{fig:dispersions} shows an isosurface of the electron density for the first singlet ($\text{S}_1$, blue, panel \textbf{a}) and triplet ($\text{T}_1$, green, panel \textbf{b}) exciton, for a hole fixed at
the center of the visualized region. As shown previously~\cite{Cudazzo2015,Refaely-Abramson2017,alve+20prb}, the singlet is significantly more
delocalized than the triplet, which results in bands that are more dispersive in reciprocal space~\cite{Cudazzo2015,Lettmann2021}, as shown in Fig.\,\ref{fig:dispersions}c. We plot the exciton energies along
the path $\Gamma \rightarrow X$ in the Brillouin zone, corresponding to the dominant packing direction of the pentacene crystal. Table\,\ref{table:widths} summarizes the bandwidth $W=\Omega(X)-\Omega(\Gamma)$ of the two excitons, as well as the width $\Delta=\Omega(\mathbf{Q}=0.4\hspace{0.1cm}\text{\AA}^{-1})-\Omega(\mathbf{Q}=0.1\hspace{0.1cm}\text{\AA}^{-1})$, the values of the 
exciton momentum
chosen to accommodate comparison to recent experiments~\cite{Graf2022}. We see from our static calculations that
the singlet bandwidth is more than twice that of the triplet.

%

We now include the effect of phonons
on the exciton band structures along $\Gamma \rightarrow X$ at $100$\,K and $300$\,K, within the harmonic and anharmonic distributions, and visualize the results in Fig.\,\ref{fig:dispersions}c
when including anharmonic effects.
There are two broad categories of phonons
in molecular crystals, corresponding to low-frequency intermolecular and high-frequency intramolecular motions, visualized in
Fig.\,\ref{fig:dispersions}d. While the former are predominantly activated when going from $100$\,K to $300$\,K, the latter have significant zero-point energies $\hbar \omega/2$. Including $100$\,K phonon effects red-shifts
both singlet and triplet exciton energies and flattens their dispersions, as shown in Fig.\,\ref{fig:dispersions}c and Table\,\ref{table:widths}. This effect
is larger for the triplet, which is more localized and therefore more impacted by high-frequency intra-molecular modes. However, increasing the temperature to $300$\,K has
no effect on the triplet, since there are negligible additional contributions from intramolecular modes at these temperatures and the modulations of intermolecular distances by lower-frequency phonons hardly affect this localized state. In contrast, the
delocalized singlet red-shifts further, and its dispersion
flattens by an additional $18$\,meV. Our results 
for the singlet width $\Delta$ at $100$\,K are in excellent agreement with recent experiments~\cite{Graf2022}, as summarized in Table\,\ref{table:widths}. 
Our predicted decrease of the singlet
width $\Delta$ by $13$\,meV when increasing the temperature from $100$\,K to $300$\,K underestimates the experimental decrease of $21$\,meV, largely
due to ignoring thermal expansion in our calculation,
which reduces $\Delta$ by a further $6$\,meV within this temperature range, see Supplemental Material~\cite{supp} Section\,S2.  Interestingly, we see
in Table\,\ref{table:widths} that the harmonic
approximation predicts an \emph{increase} of the singlet bandwidth with increasing temperature, contrary to our calculations including anharmonic effects using PIMD and to experiment, a point that we return to below.

\begin{figure}[t]
    \centering
    \includegraphics[width=0.8\linewidth]{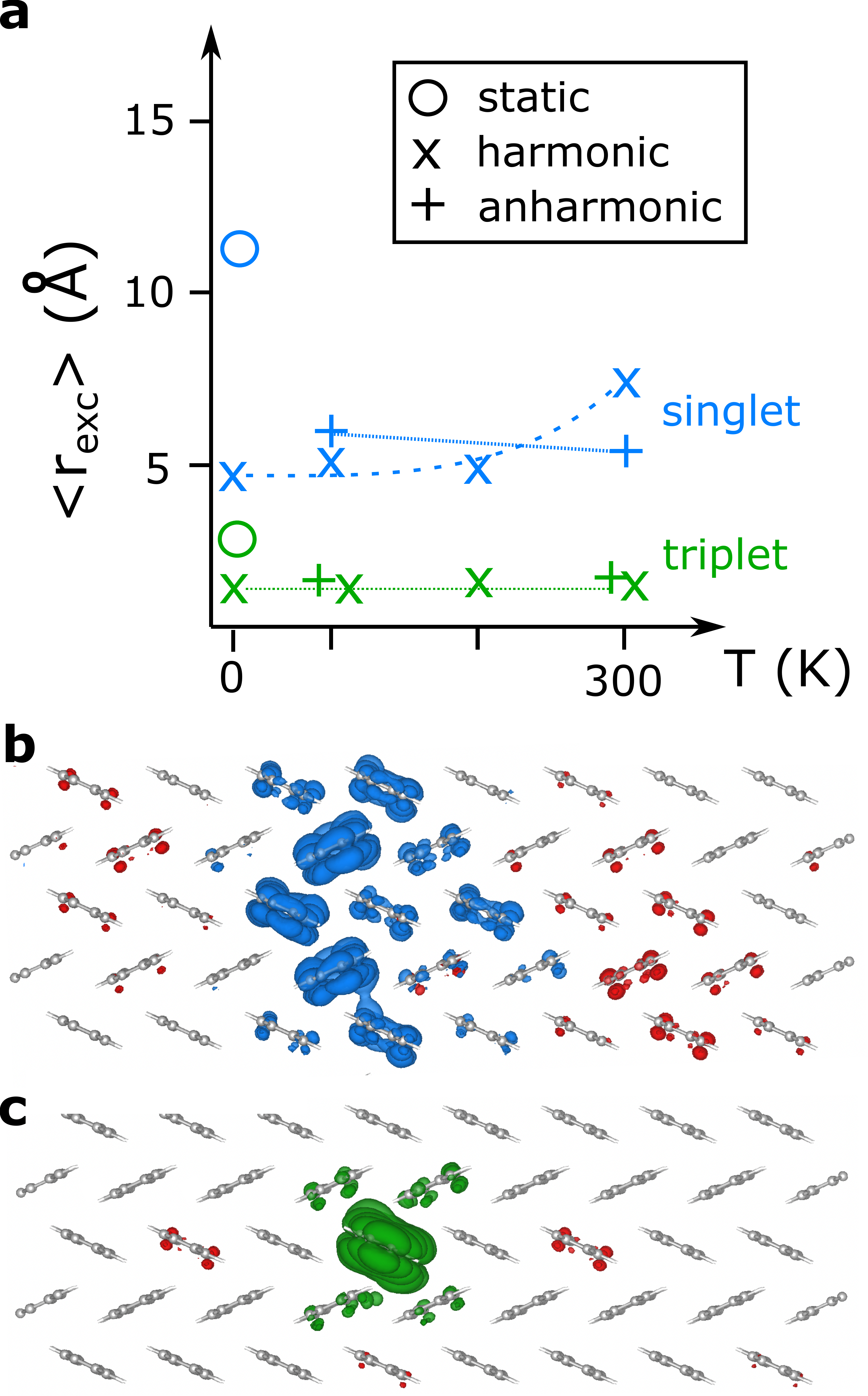}
    \caption{Singlet (blue) and triplet (green) exciton radii within the different cases and temperatures (panel \textbf{a}). Representative configuration showing electronic isosurfaces for fixed hole positions, indicating localization of the singlet (triplet) at $0$\,K towards the region in blue (green), shown in panel \textbf{b} (panel \textbf{c}).
    Red represents electronic wavefunction amplitude that disappears in the presence of phonons.}
    \label{fig:exciton_localisation}
\end{figure}

The changes in the width of the exciton dispersions suggest phonon-induced
modulations of real-space exciton properties, which are zero-point dominated for the
triplet, and which have significant temperature dependence for the singlet. We highlight the connection between the dispersion modulations and real-space exciton properties by computing vibrational averages of the exciton radii at a range of temperatures. 
The results are presented in Fig.\,\ref{fig:exciton_localisation} for the singlet (blue) and triplet (green) within the harmonic approximation and including anharmonic effects. 
Let us first comment on the harmonic case. Compared to the static limit (circles), the radii in the
presence of phonons at $0$\,K are renormalized by more than a factor of two. For the singlet, the static value of $11.2$\,\AA\hspace{0.15cm}for its radius reduces to $4.9$\,\AA, while
the static triplet radius of $2.7$\,\AA\hspace{0.1cm} reduces to $1.2$\,\AA. To visualize this we present in Fig.\,\ref{fig:exciton_localisation}b and Fig.\,\ref{fig:exciton_localisation}c differential plots for isosurfaces of the electron density once a hole is placed at a high-probability position in the unit cell. Specifically, we plot the difference between the electronic density of the case without phonons and that of a typical
atomic configuration at $0$\,K. Red indicates amplitude
vanishing due to phonons, while blue and
green indicate areas where the singlet and triplet wavefunction respectively gain amplitude, demonstrating their tendency to localize. 

When increasing the temperature to $300$\,K within the harmonic approximation there is no change to the triplet exciton radius, in agreement with our expectation of the effect of phonons on the triplet exciton dispersion. The singlet however exhibits delocalization, with its radius increasing substantially to the average value of $6.96$\,\AA, consistent with the increase of the singlet bandwidth with temperature in the harmonic case. Upon including anharmonic effects, triplet radii agree
with the harmonic case; however, for the singlet the results are qualitatively different, and we recover the expected
behavior of decreasing singlet radius with increasing temperature. All vibrational averages and errors for the exciton radii are given in Section\,S7 of the Supplemental Material~\cite{supp}. 

\begin{figure}[t]
    \centering
    \includegraphics[width=0.9\linewidth]{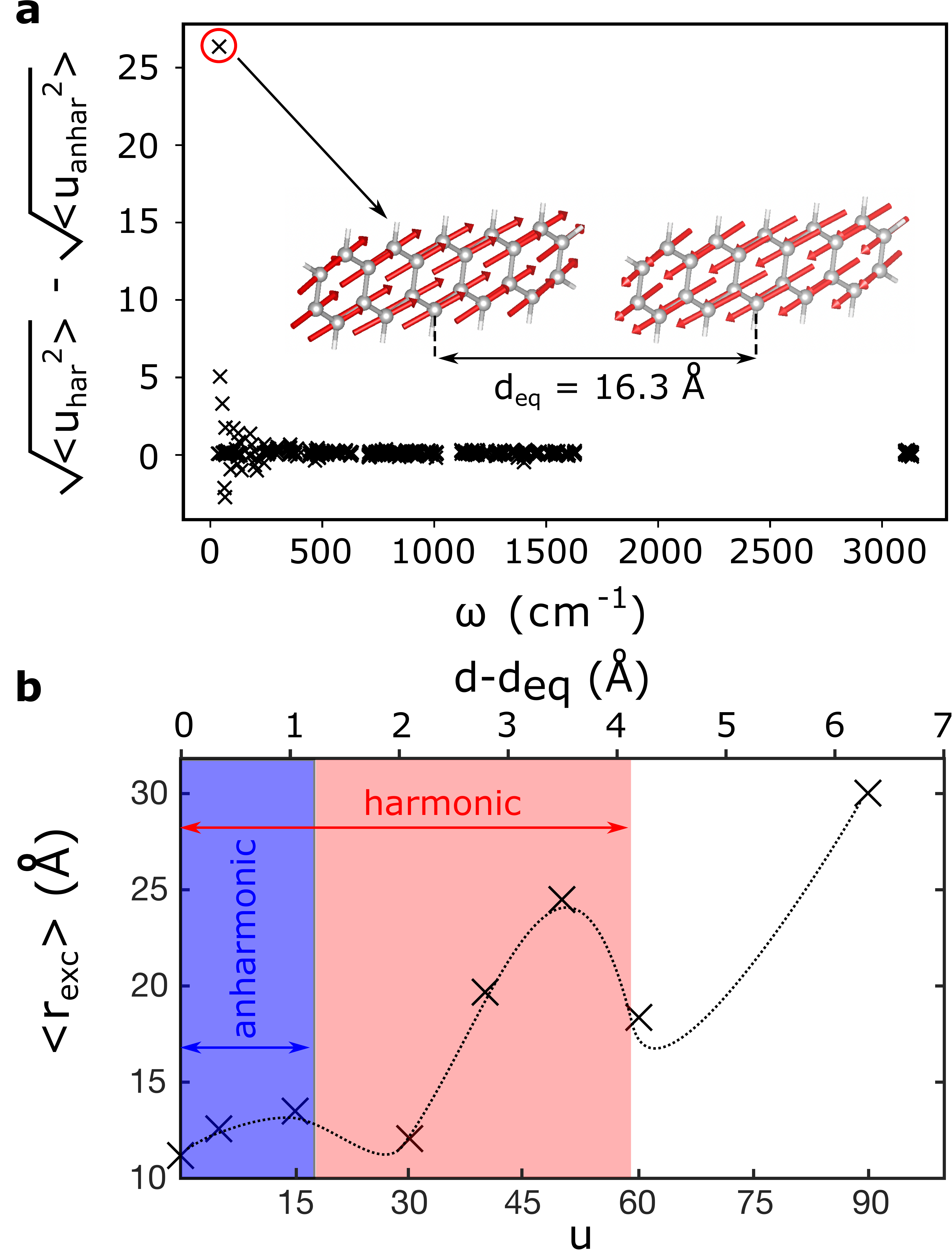}
    \caption{The difference between the RMS displacement of phonons in the harmonic and anharmonic distributions of pentacene (panel \textbf{a}). Singlet exciton radii (panel \textbf{b}) along the highly anharmonic phonon shown in panel \textbf{a}. Phonon displacements $u$ are given in units of their zero-point width $1/\sqrt{2\omega_{\boldsymbol{q}\nu}}$~\cite{Monserrat2018}. The dotted line in \textbf{b} is a guide to the eye.
    }
    \label{fig:rexc_vs_u}
\end{figure}

The discrepancy between the harmonic and anharmonic
cases is due to configurations with
highly delocalized excitons within the harmonic approximation, with radii as large as $31$\,\AA \hspace{0.1cm}at $300$\,K. Such configurations are shown in Supplementary Material~\cite{supp} Section\,S5, and their inclusion in the thermal averages of Eq.\,\ref{eq:har_band_gap} for the radii leads to
the observed temperature-induced increase of $\left\langle r_{\text{exc}} \right\rangle$ in Fig.\,\ref{fig:exciton_localisation}a.
To understand why such configurations are not present within the anharmonic case, we plot in Fig.\,\ref{fig:rexc_vs_u}a the difference between the phonon root mean squared displacement $\sqrt{\left\langle u^2 \right\rangle}$ of the two distributions at $300$\,K. We find that a
low-frequency acoustic mode, corresponding to a sliding
along the z-axis of adjacent pentacene molecules, is significantly over-displaced
in the harmonic case at $\boldsymbol{q}=X$. Anharmonic terms alter the PES associated with this phonon, limiting its average amplitude at room
temperature, as shown in Supplementary Material~\cite{supp} Fig.\,S3, in agreement with known cases where the harmonic approximation breaks down in molecular crystals~\cite{Alvertis2022,Rossi2016,Fetherolf2022}. We confirm that the over-displacement of
this phonon within the harmonic approximation leads to the temperature-induced singlet delocalization observed in Fig.\,\ref{fig:exciton_localisation}a, by computing the singlet radius as a function of amplitude of this mode, as visualized in Fig.\,\ref{fig:rexc_vs_u}b. The blue and red regions indicate the maximum range of displacements which are accessible within the anharmonic and harmonic distributions respectively, due to thermal excitation of phonons at $300$\,K. The harmonic approximation leads to configurations with highly delocalized excitons of radii as large as $25$\,\AA. The dependence of the exciton radius on the phonon displacement is non-monotonic due to the oscillating $\pi$ orbital
overlap between neighboring pentacene molecules~\cite{Arago2016}.

While highly delocalized excitons may appear at certain nuclear configurations, anharmonicity prevents accessing these, as seen in Fig.\,\ref{fig:rexc_vs_u}b. However, such configurations could appear out of equilibrium, \emph{e.g.} due to photoexcitation, upon relaxation to the excited state PES minimum. For pentacene, the minimum of
the singlet exciton PES along the anharmonic acoustic mode lies far from the
`delocalized' region of Fig.\,\ref{fig:rexc_vs_u}b (see Supplemental Material~\cite{supp} Section\,S6), it is thus unlikely that for this and similar systems 
transiently
delocalized excitons may
be accessed, even outside equilibrium.

\emph{Conclusions.}-- We have presented a first-principles
study of the effect of phonons on the dispersion and radii of excitons
in the prototypical molecular crystal pentacene. Zero-point nuclear motion uniformly causes
substantial localization of excitons, manifesting as a flattening of the exciton dispersion
in reciprocal space. 
Wannier-Mott-like singlet excitons also exhibit additional temperature-activated localization due to their stronger coupling to low-frequency phonons, with anharmonic effects
being critical in capturing this effect
and preventing transient exciton delocalization. 
Anharmonic low-frequency phonons are common in molecular materials~\cite{Alvertis2022} and can couple to singlets when these approach the Wannier-Mott limit, in a manner which is in turn determined by the size~\cite{alve+20prb} and packing~\cite{Sharifzadeh2015} of the molecular building blocks. 
Our work lays foundations for a deep understanding and controlled enhancement of exciton transport in molecular crystals, for example by suppressing anharmonicity through
chemical modifications~\cite{Asher2022}.

We thank Sivan Refaely-Abramson for useful discussions. This work was primarily supported by the
Theory FWP, which provided $GW$ and $GW$-BSE calculations and analysis of phonon effects, and the Center for Computational Study of Excited-State Phenomena in Energy Materials (C2SEPEM), which provided advanced codes, at the Lawrence Berkeley National Laboratory, funded by the U.S. Department of Energy, Office of Science, Basic Energy Sciences, Materials Sciences and Engineering Division, under Contract No. DE-AC02-05CH11231. SS acknowledges funding from the U.S. National Science Foundation (NSF) under grant number DMR-1847774. Computational resources were provided by the National Energy Research Scientific Computing Center (NERSC).

\textbf{}
\bibliography{references}

\end{document}


\title{\textbf{\Large{Supplemental Material: Phonon-induced localization of excitons in molecular crystals from first principles}}}

\author{Antonios M. Alvertis}
\email{amalvertis@lbl.gov}
\affiliation{Materials Sciences Division, Lawrence Berkeley National Laboratory, Berkeley, California 94720, USA}
\affiliation{Department of Physics, University of California Berkeley, Berkeley, United States}
\author{Jonah B. Haber}
\affiliation{Department of Physics, University of California Berkeley, Berkeley, United States}
\author{Edgar A. Engel}
\affiliation{Cavendish Laboratory, University of Cambridge, J.\,J.\,Thomson Avenue, Cambridge CB3 0HE, United Kingdom}
\author{Sahar Sharifzadeh}
\affiliation{Division of Materials Science and Engineering, Boston University, United States}
\affiliation{Department of Electrical and Computer Engineering, Boston University, United States}
\author{Jeffrey B. Neaton}
\email{jbneaton@lbl.gov}
\affiliation{Materials Sciences Division, Lawrence Berkeley National Laboratory, Berkeley, California 94720, USA}
\affiliation{Department of Physics, University of California Berkeley, Berkeley, United States}
\affiliation{Kavli Energy NanoScience Institute at Berkeley, Berkeley, United States}


\maketitle

\tableofcontents

\clearpage
\section{Computational details}

\subsection{DFT calculations}

First-principles energies, forces, band gaps and wavefunctions are computed using the Quantum Espresso~\cite{QE} DFT code with the semi-local PBE exchange-correlation functional~\cite{pbe} and a plane-wave energy cut-off of 60\,Rydberg for the wavefunction. For the unit cell of
pentacene we start from the experimental structure PENCEN08 as in the Cambridge Crystallographic Database~\cite{pent}, and relax the internal coordinates while leaving the volume fixed and using the Tkatchenko-Scheffler (TS) dispersion correction~\cite{ts}. For the geometry optimization we employ a $4\times4\times2$ Monkhorst-Pack $\mathbf{k}$-point grid~\cite{monkhorst_1976}.

\subsection{$GW$-BSE calculations}
We employ the one-shot $G_oW_o$ approximation for the quasiparticle
properties of pentacene, as implemented in the BerkeleyGW code~\cite{Deslippe2012}. We use a $4\times4\times2$ k-grid, $400$ bands and a $7$\,Ry plane wave cutoff to calculate the
dielectric screening. For exciton calculations, we
construct the electron-hole kernel on a $4\times4\times2$ grid using $4$ valence and $4$ conduction states, and then interpolate on a $8\times8\times4$ k-grid with the
same number of bands. This set of parameters has been
shown to give converged results in previous computational
studies of pentacene~\cite{Refaely-Abramson2017,alve+20prb}. When studying $2\times1\times1$ supercells of pentacene (see following subsection on the Monte Carlo sampling of vibrational averages), we use a half the
number of k-points in the x direction in all cases ($2\times4\times2$ to calculate the dielectric screening and $4\times8\times4$ to interpolate the BSE kernel), and double the
number of bands for the dielectric screening ($800$ bands) and exciton calculations ($8$ valence and $8$ conduction bands).

\subsection{Phonon calculations and Monte Carlo sampling of vibrational averages}
\label{monte_carlo}

\begin{figure}[t]
    \centering
    \includegraphics[width=0.7\linewidth]{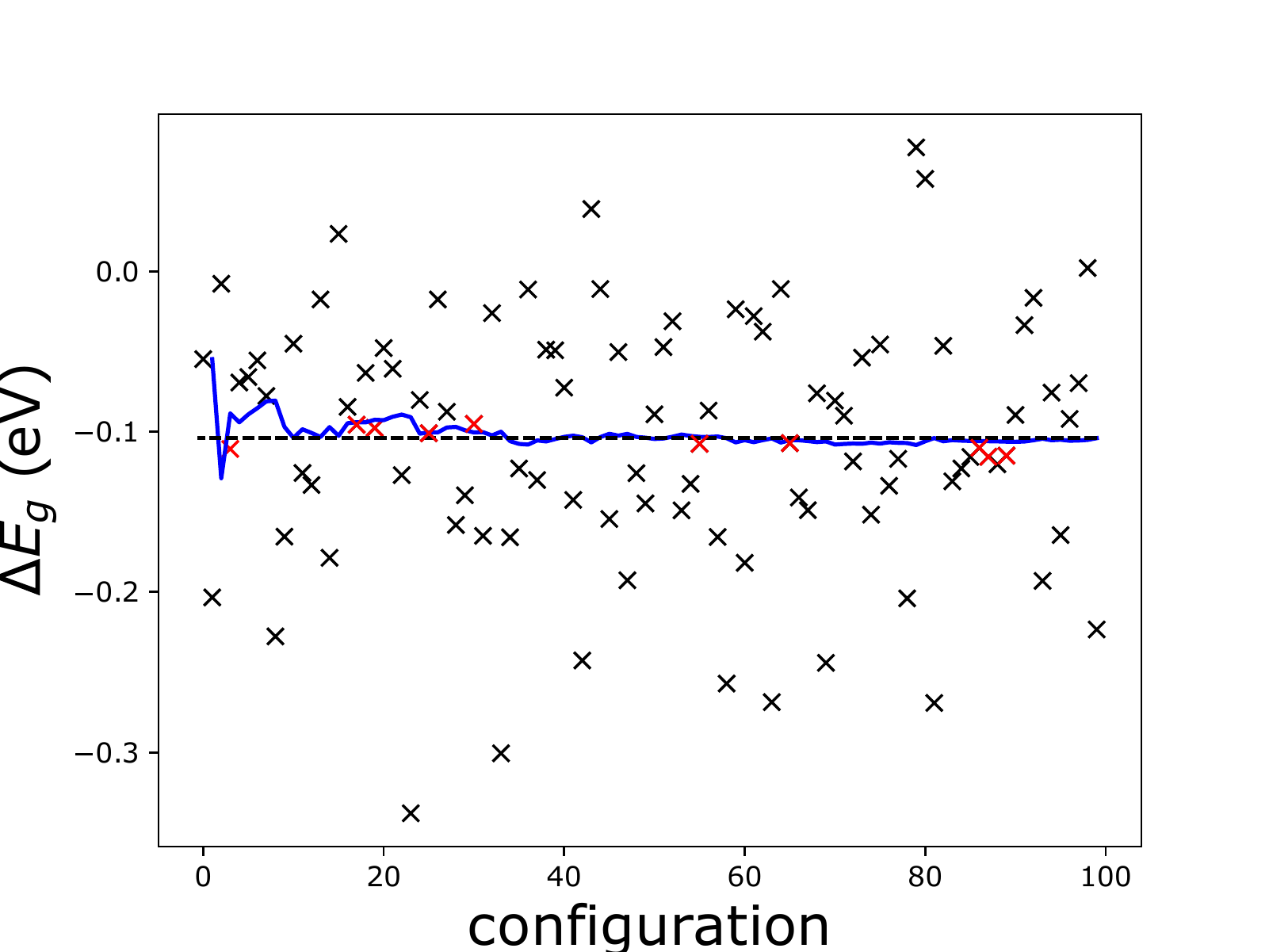}
    \caption{Convergence of the DFT band gap renormalization $\Delta E_g$ as a function of the number of configurations used in the Monte Carlo sampling, for the example of using a $2\times1\times1$ supercell at $T=100$\,K. The value of the gap renormalization for each configuration is denoted in black crosses, while the blue line is the running average of the band gap renormalization due to phonons. The dashed line is the final computed average value for the band gap renormalization, while the red crosses indicate the ten configurations for which we run $GW$-BSE calculations to obtain their exciton properties.}
    \label{fig:convergence_MC}
\end{figure}

We include the contribution from lattice dynamics at temperature $T$ to an observable $\mathcal{O}$ by means of the quantum mechanical expectation value
\begin{equation}
    \mathcal{O}(T)=\frac{1}{\mathcal{Z}}\sum_{\mathbf{s}}\langle\chi_{\mathbf{s}}(\mathbf{u})|\mathcal{O}(\mathbf{u})|\chi_{\mathbf{s}}(\mathbf{u})\rangle e^{-E_{\mathbf{s}}/k_{\mathrm{B}}T}, \label{eq:exp_value}
\end{equation}
where $|\chi_{\mathbf{s}}\rangle$ is the harmonic vibrational wavefunction in state $\mathbf{s}$ with energy $E_{\mathbf{s}}$, $\mathcal{Z}=\sum_{\mathbf{s}}e^{-E_{\mathbf{s}}/k_{\mathrm{B}}T}$ is the partition function, and $\mathbf{u}=\{u_{\mathbf{q}\nu}\}$ is a collective coordinate that includes the amplitudes of all normal modes of vibration in the system labeled by the phonon wave vector $\mathbf{q}$ and the phonon branch $\nu$. 

Substituting the harmonic vibrational wavefunction, the above expectation value can be rewritten

\begin{equation}
    \mathcal{O}(T)=\int d\mathbf{u}|\Phi(\mathbf{u};T)|^2\mathcal{O}(\mathbf{u}), \label{eq:harm_exp_value}
\end{equation}
where: 
\begin{equation}
    |\Phi(\mathbf{u};T)|^2 = \prod_{\mathbf{q},\nu}(2\pi \sigma^2_{\mathbf{q}\nu}(T))^{-1/2}\exp{\left(-\frac{u_{\mathbf{q}\nu}^2}{2\sigma^2_{\mathbf{q}\nu}(T)}\right)},
    \label{eq:harm_density}
\end{equation}
the harmonic density at temperature $T$, which is a product of Gaussian functions of width:
\begin{equation}
    \sigma^2_{\mathbf{q}\nu}(T) = \frac{1}{2\omega_{\mathbf{q}\nu}}\cdot \coth{\left(\frac{\omega_{\mathbf{q}\nu}}{2k_BT}\right)}.
    \label{eq:Gaussian_width}
\end{equation}

We evaluate Eq.\,(\ref{eq:harm_exp_value}) by generating stochastic samples distributed according to the harmonic vibrational ensemble, calculating the observable of interest at each configuration, and averaging over all configurations. 
To sample the single-particle DFT electronic band gap  we generate $100$ configurations, which are sufficient for convergence, as demonstrated in Fig.\,\ref{fig:convergence_MC}. We obtain the band gap correction for each of these configurations at temperatures of $T=0$\,K, $T=100$\,K, $T=200$\,K and $T=300$\,K. 
We then apply $GW$ corrections to these DFT values. Due to the large computational cost of these calculations, we only perform $GW$ calculations on the ten configurations whose single-particle DFT band gap value is closest to the calculated average band gap for each temperature, as also shown in Fig.\,\ref{fig:convergence_MC}. This correlated sampling strategy between DFT and $GW$ has been shown to be accurate in pentacene~\cite{alve+20prb} and other systems~\cite{Monserrat2016}. Having calculated the effects of electron-phonon coupling on the quasiparticle band gap, we solve the Bethe-Salpeter equation for the same ten configurations at the various temperatures, using the parameters of the previous section and computing exciton energies (at finite exciton momenta)
and exciton radii.

\begin{figure}[t]
    \centering
    \includegraphics[width=0.7\linewidth]{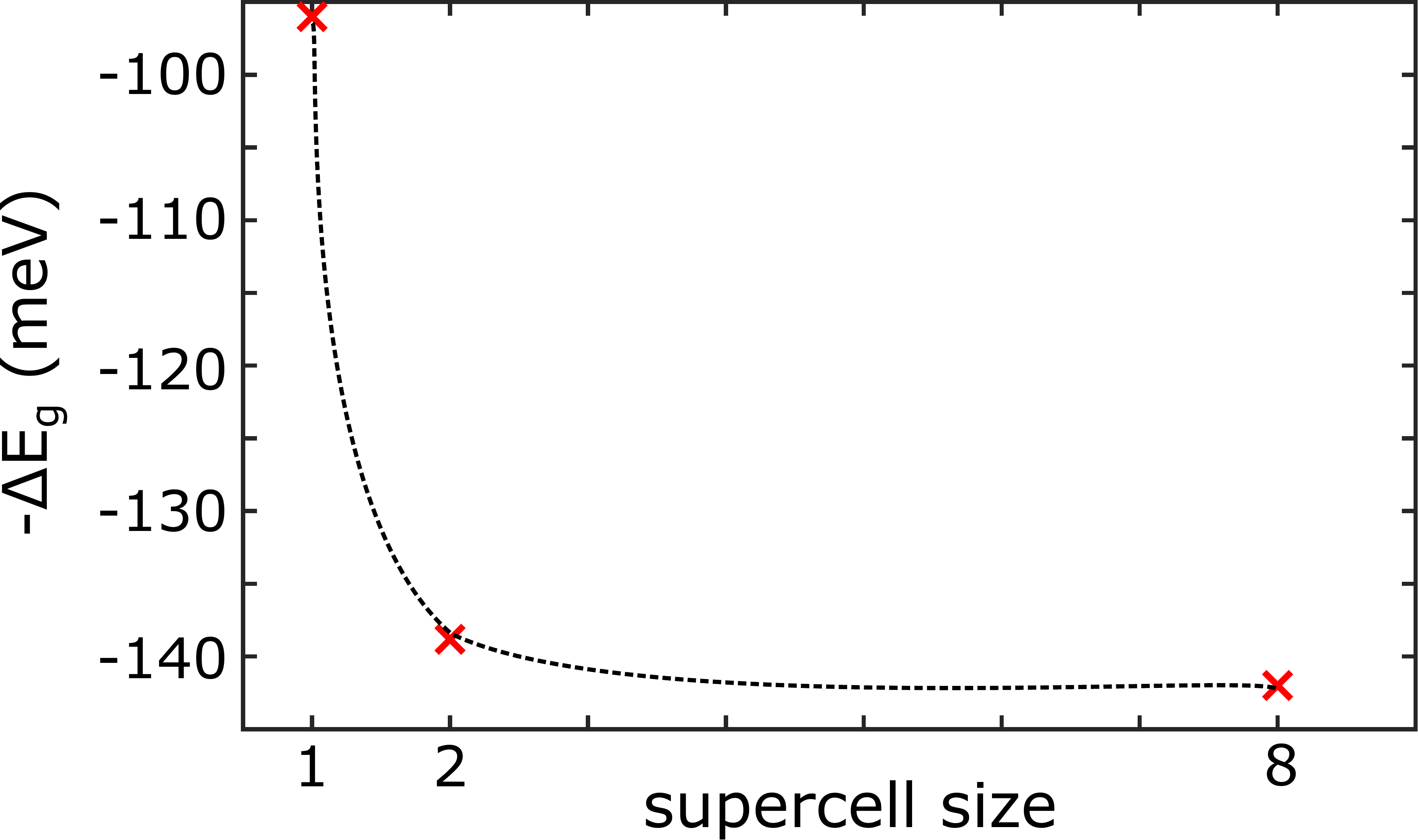}
   \caption{Convergence of the pentacene DFT band gap renormalization $\Delta E_g$ at $T=100$\,K, as a function of the  supercell size. The dashed line is given as a guide to the eye.}
    \label{fig:convergence_supercell}
\end{figure}

The sampling of the expectation value of equation\,\ref{eq:harm_exp_value} becomes increasingly accurate with the inclusion of more $\mathbf{q}$-points in the Brillouin zone. Within the finite differences approach for phonon calculations and the expectation values of observables at finite temperatures, $\mathbf{q}$-points are described using commensurate supercells. For pentacene a $2\times1\times1$ (size 2) supercell (four pentacene molecules) is $98$\% converged with respect to a $2\times2\times2$ (size 8) supercell (sixteen molecules) for the band gap zero-point renormalization ($-139$\,meV and $-142$\,meV respectively), as seen in Fig.\,\ref{fig:convergence_supercell} where we plot 
the convergence of the pentacene band gap renormalization at $100$\,K, as obtained from sampling within the anharmonic distribution.
Therefore a $2\times1\times1$ supercell offers a good
balance between computational cost for the $GW$-BSE calculations, and accuracy. 

\subsection{Machine learning potential and path-integral molecular dynamics}

The details of the construction of the machine learning potential for pentacene (and other acene crystals), as well as the procedure for obtaining
trajectories within path-integral molecular
dynamics, have been described elsewhere~\cite{Alvertis2022}. Here we include
some key points and refer the reader interested
in a more in-depth discussion to Ref.~\cite{Alvertis2022}. 

A machine learning potential describing the dynamics of the
acene series of molecular crystals was trained on a set of training data including the total energies of $4862$ configurations of naphthalene, anthracene, tetracene and pentacene, obtained from the harmonic distributions at 0\,K, 150\,K, and 300\,K. From this set, 400 validation and 400 test configurations were drawn randomly. For tuning the ML potential architecture and training procedure, the training data was sparsified by farthest-point-sampling (FPS)~\cite{fps1, fps2},  retaining the 1000 most structurally distinct training configurations. 
This FPS was performed on the basis of Euclidean distances between configurations described in terms of their smooth overlap of atomic positions (SOAP) powerspectra~\cite{soap}, using the radially-scaled implementation~\cite{rs_soap} with a radial and angular basis of 12 and nine functions, respectively, a cut-off radius of $8\Ang$, a width of $0.275\Ang$ for the Gaussian densities associated with the atomic positions, and a scaling onset and exponent of $2.5\Ang$ and $4.5$, respectively.
An ensemble of seven fully-connected, feed-forward neural networks with two hidden layers of 16 nodes each was constructed by using the N2P2 code~\cite{N2P2}. 

For the independent testset, the ML potentials reproduce the reference energies and forces with root-mean-square errors of 2.4\,meV/atom and 0.157\,meV/$\Ang$, respectively.
Crucially, this suffices to run stable path integral simulations in the constant-volume ensemble over extended simulation times, and to accurately compute the quantum-mechanical expectation values of observables within the reference first-principles thermodynamic ensemble.

Finite-temperature, quantum-mechanical thermodynamic averages of observables can be computed as averages of their values for (random) configurations drawn from PIMD simulations. 
We exploit the affordability of the ML potential to perform PIMD in the $NVT$ ensemble at temperatures of 100\,K and 300\,K, and subsequently compute observables for configurations extracted from the PIMD trajectory (after equilibration) at regular intervals of 50\,fs, which ensures that these samples are  decorrelated.
The PIMD simulations of at least 10\,ps were performed using the i-PI~\cite{ipi} molecular dynamics engine to drive LAMMPS~\cite{lammps} energy and force evaluations of the ML potential, a 0.25\,fs timestep, and a path integral Langevin equation thermostat~\cite{ceri+10jcp} with $\tau = 100$\,fs.

The number of replicas required for PIMD simulations is determined by the highest frequency phonon modes that are present in the system, which in our case is C--H stretching and is common among
all acene crystals. For benzene it has been shown
that 32 replicas are sufficient to convergence
electronic band gaps within $15$\,meV\cite{Kapil2022,Alvertis2022}, and the
same value has therefore been employed for pentacene. 

For obtaining the exciton properties (energies and radii) within the anharmonic distributions we follow a tactic similar to the correlated sampling described in the case of the harmonic Monte Carlo sampling. We first draw $100$ configurations at which we compute the DFT band gap and then we rank 
the configurations based on their band gap proximity to the computed mean value. We then perform $GW$-BSE calculations on the top 10 configurations among these.

\subsection{Calculation of exciton radii}

As discussed in the main text, we compute the electron-hole correlation function as defined in Ref.~\cite{Sharifzadeh2013}, namely

\begin{equation}
    \label{eq:correlation_function}
    F_S(\boldsymbol{r})=\int_{V}d^3\boldsymbol{r}_h |\psi^{\boldsymbol{Q}=0}_S(\boldsymbol{r}_e=\boldsymbol{r}_h+\boldsymbol{r},\boldsymbol{r}_h)|^2,
\end{equation}
where $V$ the volume of the primitive cell. $F_S(\boldsymbol{r})$ describes the probability of finding the electron-hole pair at a vector $\boldsymbol{r}=\boldsymbol{r}_e-\boldsymbol{r}_h$, and
is computed as a discrete sum over hole positions. We note here that
even if we only integrate over the volume of the primitive cell, the exciton wavefunction can delocalize over the whole supercell used in the Bethe-Salpeter calculation, \emph{i.e.}, $8\times8\times4$ or $4\times8\times4$ for pentacene, as discussed below. Even if the hole was moved outside
the primitive cell, we would simply obtain a shifted exciton wavefunction.

For pentacene, it was found in Ref.~\cite{Sharifzadeh2013}
 that the average electron-hole distance of the correlation function is converged, and that its envelope produces a smooth function for $88$ high-probability hole
positions in the unit cell, corresponding
to two hole positions per carbon atom, at $\pm0.5\AA$ above and below the plane
of the molecule for each atom (effectively sampling the C $p_z$ orbitals). For each
sampled hole, the correlation function
is computed on an $8\times8\times4$
supercell of pentacene, which is necessary for convergence of the
exciton wavefunctions. For the $2\times1\times1$ cell of pentacene, which includes the effects of phonons (see section\,\ref{monte_carlo}), the real-space supercell used for $F_S(\boldsymbol{r})$ is reduced to $4\times8\times4$, since every cell along the x direction already contains two unit cells. Moreover the number of carbon atoms per cell doubles, and so we sample at $88\cdot 2=176$ hole positions in this case.

Having computed the electron-hole correlation function for a given atomic configuration $\boldsymbol{u}$, the  exciton radius is computed as

\begin{equation}
r_{\text{exc}}(\boldsymbol{u})=\int d|\boldsymbol{r}|F_S(|\boldsymbol{r}|)|\boldsymbol{r}|.
\end{equation}
To compute the vibrationally renormalized
exciton radii at a temperature $T$, we apply equations 5 and 7 of the main manuscript with $\mathcal{O}=r_{\text{exc}}$. In the example case where we make the harmonic approximation this results in

\begin{equation}
    \left\langle r_{\text{exc}}(T)\right\rangle=\int d\mathbf{u}|\Phi(\mathbf{u};T)|^2r_{\text{exc}}(\mathbf{u}), \label{eq:harm_exp_value_rexc}
\end{equation}
where $|\Phi(\mathbf{u};T)|^2$ the
harmonic density function, as discussed in section\,\ref{monte_carlo} above.

\section{Effect of thermal expansion}

To study the effect of thermal
expansion on the exciton dispersion
of pentacene, we perform $GW$-BSE 
calculations on two pentacene structures
deposited in the Cambridge Crystallographic Database,
which have been obtained through X-ray diffraction measurements
at temperatures within the range of interest of $100-300$\,K. These are
the structures PENCEN06 and PENCEN07~\cite{pent}, obtained at $120$\,K and $293$\,K respectively. We use the experimental crystal structures without
any optimization of the internal coordinates. 
The absolute value of the singlet exciton energy
and width are known to be very sensitive to coordinate optimization and the precise level of
theory employed to perform this optimization~\cite{Rangel2016}. Therefore here we will focus on differences between energies and dispersion widths of the two experimental structures, without attempting a direct comparison
to values obtained for the optimized structure
necessary to perform phonon calculations. In principle phonons and thermal expansion need to be
accounted for concurrently, however the large number
of degrees of freedom in molecular crystals constitute such an analysis extremely challenging. 

The dispersion width $W = E(X)-E(\Gamma)$ is found to be
$11$\,meV smaller in the high-temperature phase. 
Given the linear character of the static exciton
dispersion of pentacene, we can
estimate the bandwidth $\Delta=\Omega(\mathbf{Q}=0.4\hspace{0.1cm}\text{\AA}^{-1})-\Omega(\mathbf{Q}=0.1\hspace{0.1cm}\text{\AA}^{-1})$, and
find that $\Delta$ for the singlet
exciton shrinks by $6$\,meV due to thermal expansion within this
range of temperatures. There is therefore no competition between thermal expansion and exciton-phonon coupling in terms of their effect on the exciton dispersion. Moreover, this flattening of the exciton dispersion caused by thermal expansion
will bring the predicted value of $\Delta=30$\,meV for the
width at $300$\,K and when including anharmonic effects, even closer to the experimental value of $23$\,meV at the same temperature (Table I, main manuscript).

\section{Contributions of $\Gamma$ and $X$ phonons}

\subsection{Exciton radii}
The values for the vibrationally renormalized exciton radii given in the
main manuscript, in both the harmonic
and anharmonic cases, include the effects
of $\Gamma$ and $X$ phonons, \emph{i.e.}
include phonons within a $2\times1\times1$ supercell of pentacene.
As shown in Fig.\,2 of the main manuscript, the harmonic approximation
predicts an increase of the average exciton radius with increasing the temperature from $0$\,K to $300$\,K. 
Qualitatively, this is in agreement
with the case of only including the effect of $\Gamma$ phonons within the
harmonic approximation, \emph{i.e.} focusing on a single unit cell of pentacene, as seen in Table\,\ref{table:rexc_Gamma}. Quantitatively, this increase of the radius becomes more prevalent when including $X$ phonons (in the $2\times1\times1$ supercell), which is
to be expected given that anharmonic
effects are more relevant for finite
phonon wavevectors $\boldsymbol{q}$ as
discussed in the main manuscript and in
Ref.~\cite{Alvertis2022}.

\begin{table}[tb]
\centering
  \setlength{\tabcolsep}{6pt} 
\begin{tabular}{ccc}
\hline
$r_{\text{exc}}(T)$ (\AA)& $1\times1\times1$ ($\Gamma$) & $2\times1\times1$ ($\Gamma,X$)\\
\hline
$0$\,K & $4.73\pm0.35$ & $4.85\pm0.30$ \\
$300$\,K & $5.09\pm0.40$ & $6.96\pm1.25$ \\
\hline
\end{tabular}
\caption{The effect of $\Gamma$ and $X$ phonons on the vibrationally-averaged exciton radius, within the harmonic approximation. }
\label{table:rexc_Gamma}
\end{table}

\subsection{Exciton dispersion}

Tables\,\ref{table:anhar_dispersion} and \ref{table:har_dispersion} summarize the effect of $\Gamma$ phonons only on the width $W=E(X)-E(\Gamma)$
of the exciton dispersion within the anharmonic and harmonic distributions respectively. The results
given in Table\,I of the main manuscript also contain
the effect of phonons at the band-edge $X$.

For both the harmonic and anharmonic 
cases, the result for the triplet is
the same: the width of the exciton
band narrows entirely due to zero-point
motion, and increasing the temperature
to $300$\,K has no effect on it. For
the singlet, while we find that including anharmonic effects at $100$\,K leads to a greater band-narrowing compared to the harmonic case (at $0$\,K), increasing the temperature to $300$\,K leads to a further reduction of the bandwidth
by $8$\,meV, compared to the $12$\,meV of the harmonic case. For the exact
values of the exciton energies please
refer to the tables of section\,S7. It is also worth
noting that in this case of including $\Gamma$ phonons only, the harmonic approximation does not
show the unphysical increase of the dispersion width
with increasing temperature, due to not including
the highly anharmonic acoustic phonon at $\mathbf{q}=X$, see also section\,S4.

\begin{table}[tb]
\centering
  \setlength{\tabcolsep}{6pt} 
\begin{tabular}{ccc}
\hline
 & $W(\text{S}_1)$ (meV) & $W(\text{T}_1)$ (meV) \\
\hline
static & $110$ & $52$ \\
$100$\,K & $67$ & $18$ \\
$300$\,K & $59$ & $19$ \\
\hline
\end{tabular}
\caption{The effect of $\Gamma$ phonons on the dispersion width $W$ for the first singlet and triplet excitons of pentacene, including anharmonic effects.}
\label{table:anhar_dispersion}
\end{table}

\begin{table}[tb]
\centering
  \setlength{\tabcolsep}{6pt} 
\begin{tabular}{ccc}
\hline
 & $W(\text{S}_1)$ (meV) & $W(\text{T}_1)$ (meV) \\
\hline
static & $110$ & $52$ \\
$0$\,K & $82$ & $20$ \\
$300$\,K & $70$ & $19$ \\
\hline
\end{tabular}
\caption{The effect of $\Gamma$ phonons on the dispersion width $W$ for the first singlet and triplet excitons of pentacene, within the harmonic approximation.}
\label{table:har_dispersion}
\end{table}

\section{Anharmonic potential energy surfaces}

\begin{figure}[t]
    \centering
    \includegraphics[width=0.7\linewidth]{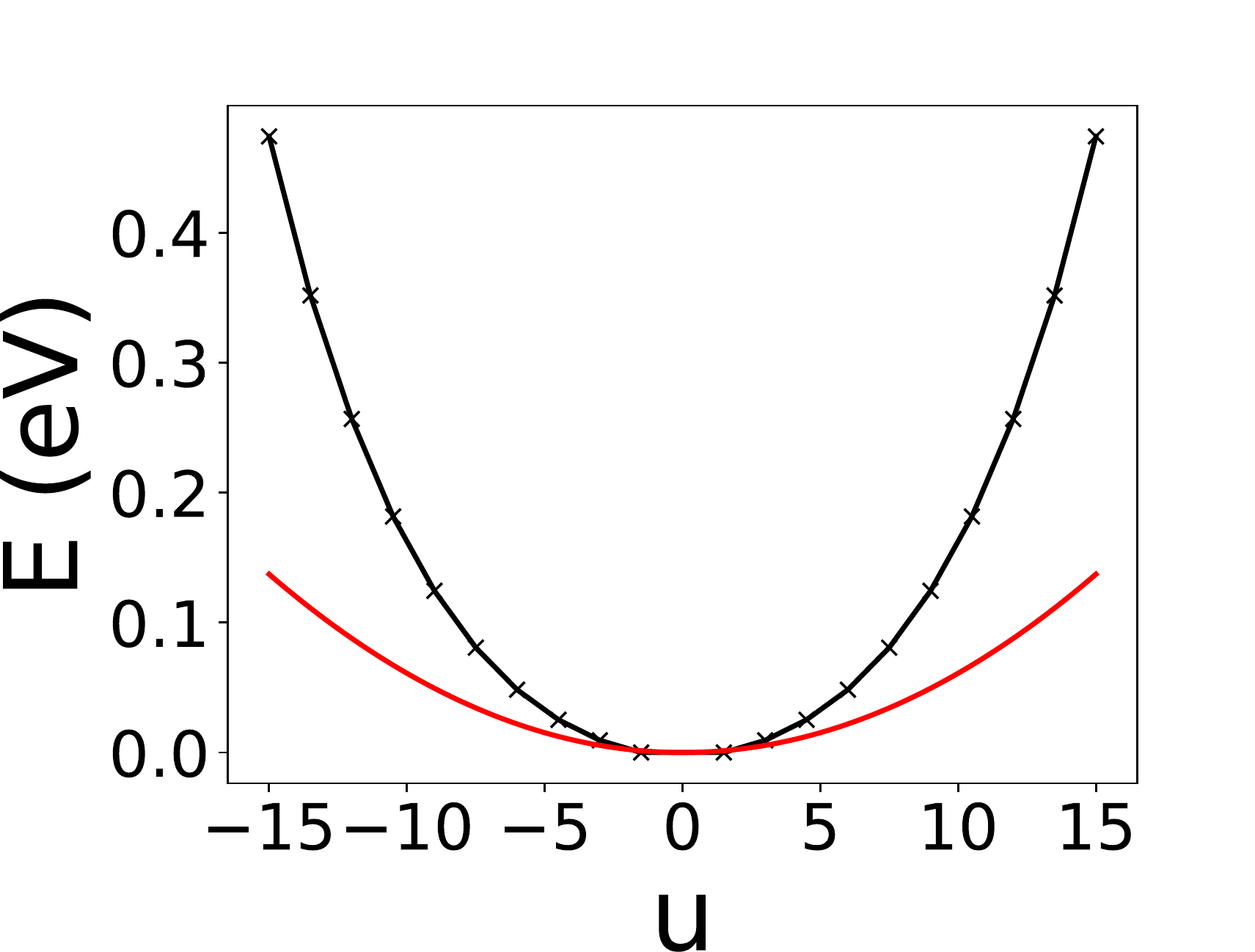}
    \caption{Comparison between the harmonic (red) and anharmonic (black) potential energy surfaces of the acoustic mode with frequency $\omega=40$\,$\text{cm}^{-1}$ at $\boldsymbol{q}=X$. The phonon displacement $u$ is given in units of the zero-point width $\frac{1}{\sqrt{2\omega}}$.}
    \label{fig:anharmonic_PES}
\end{figure}

Fig.\,\ref{fig:anharmonic_PES} visualizes
the potential energy surface of the
anharmonic acoustic mode of pentacene
discussed in the main manuscript, at $\boldsymbol{q}=X$, where it has a frequency of $\omega=40$\,$\text{cm}^{-1}$. In red
we show the harmonic potential energy
surface as predicted by the relationship
$E=\frac{1}{2}\omega^2u^2$, while the black crosses indicate the total energy
of the system (with respect to that of
the optimized geometry) upon explicitly
displacing along this phonon and performing DFT calculations at different
displacements $u$. While for small values of $u$ the two results coincide, they
quickly start to diverge, and anharmonicity provides an energetic barrier which prevents the over-displacement of pentacene along
this mode, as permitted within the harmonic approximation.

\section{Highly delocalized excitons within the
harmonic approximation}

\begin{figure}[t]
    \centering
    \includegraphics[width=0.5\linewidth]{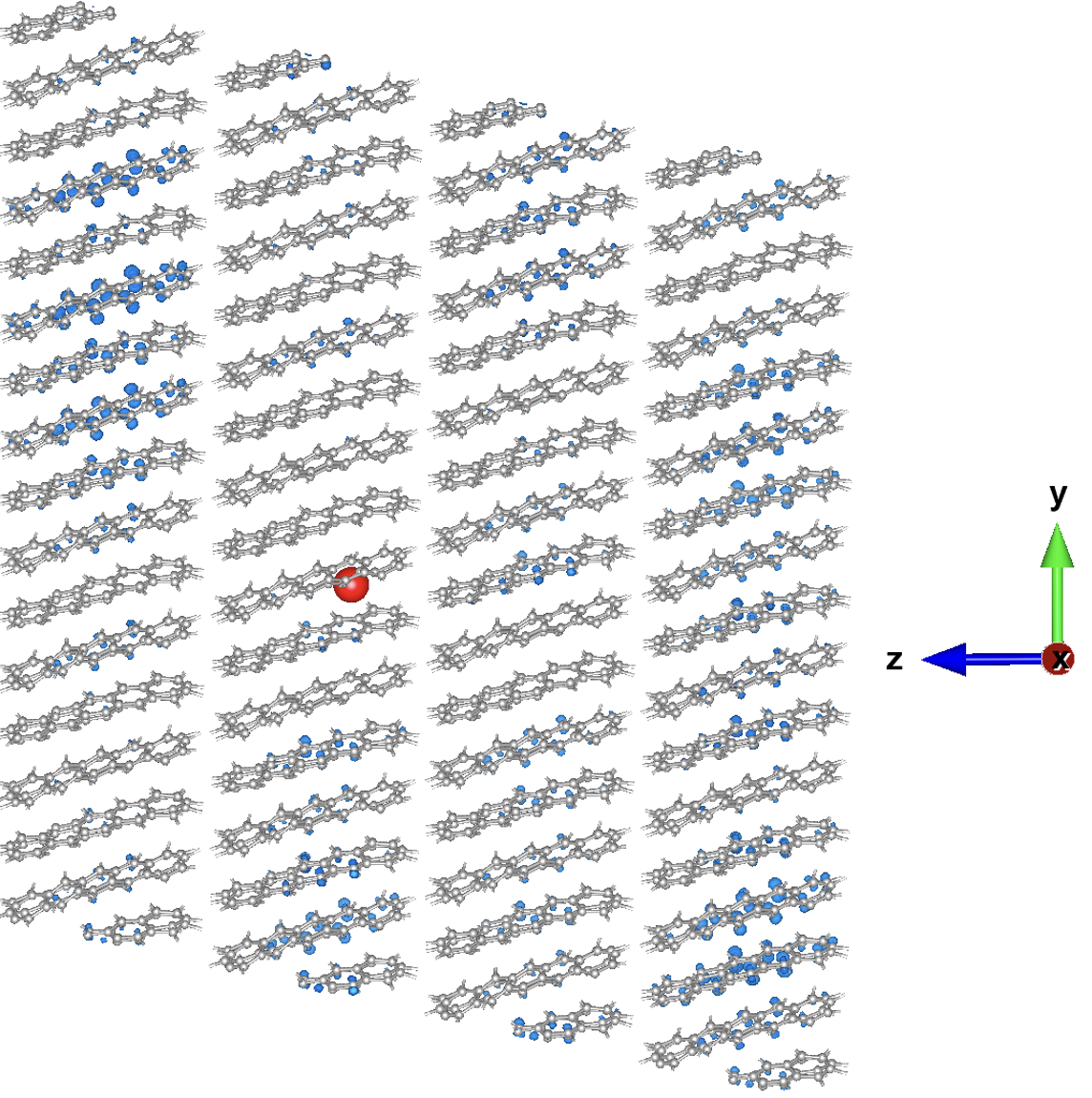}
    \caption{Isosurface of the electron distribution (blue) for a highly delocalized pentacene singlet exciton, for a hole placed at the position in red. This exciton with a radius of $31$\,\AA \hspace{0.1cm} appears within the harmonic approximation for a $2\times 1 \times 1$, as a result of thermal phonon excitation at $300$\,K.}
    \label{fig:delocalized_exciton}
\end{figure}

As discussed in the main manuscript, the
thermal excitation of phonons within the harmonic
approximation leads to configurations with highly
delocalized excitons. The most prominent examples
appear when the anharmonic acoustic phonon
of pentacene, corresponding to a sliding of adjacent molecules along the z-axis, is displaced 
significantly. Let us consider the case of thermal excitation at $300$\,K for a $2\times 1 \times 1$ supercell, which describes
this phonon at $\mathbf{q}=X$. In this case, several high-probability hole positions
used to sample the electron-hole correlation function of Eq.\,\ref{eq:correlation_function}
lead to very large radii in the range of $25-30$\,\AA, for several nuclear geometries. It is worth noting that for some
nuclear configurations, it is the hole position
with the highest probability entering Eq.\,\ref{eq:correlation_function} which already 
leads to large exciton radii. As an example, we plot in Fig.\,\ref{fig:delocalized_exciton} the
delocalized exciton that arises for a configuration used in the sampling of exciton properties. Here the hole
is placed in the most likely position out of the $176$ ones used for the sampling of the electron-hole correlation function of the specific nuclear configuration, resulting in an exciton radius of $31$\,\AA. Other examples include two sampled nuclear configurations where the highest probability hole positions lead to excitons with radii of $23$\,\AA \hspace{0.1cm}and $25$\,\AA. It is
also emphasized that for \emph{every} nuclear configuration resulting from thermal phonon excitation at $300$\,K within the harmonic approximation we find highly probable hole positions that lead to similarly highly delocalized excitons, although in the above three examples these are not only high-probability holes, but the ones with the highest probability.

\section{Estimation of the minimum of the
singlet exciton potential energy surface}

We would like to estimate the position of the minimum of the singlet exciton potential energy surface along the acoustic phonon of pentacene, which is responsible for exciton delocalization at large displacements $u$. To do so, we have
to assume the ground and excited state potential energy surfaces to be harmonic. The ground state energy along this phonon will be
\begin{equation}
    \label{eq:GS_PES}
    E_{GS}(u)=\frac{1}{2}\omega_{GS}^2u^2,
\end{equation}
where $\omega_{GS}$ the phonon frequency in the ground state. 
Reexpressing the phonon displacement $u$ in units of the zero-point width $1/\sqrt{2\omega}$ of the harmonic distribution, this may be rewritten as

\begin{equation}
    \label{eq:GS_PES_sigma}
    E_{GS}(u)=\frac{1}{4}\omega_{GS}u^2.
\end{equation}
The excited state surface within these units is written as
\begin{equation}
    \label{eq:ES_PES_sigma}
    E_{ES}(u)=\frac{1}{4}\omega_{ES}(u-u_{ES})^2+\Delta,
\end{equation}
where we have the following unknown quantities: $u_{ES}$, which is the position at which the
excited state obtains its minimum energy, $\Delta$, which is
the total energy of the system when in the excited state and at that geometry and $\omega_{ES}$, which
is the excited state mode frequency. 

We would now like to estimate $u_{ES}$, for this purpose we perform DFT and $GW$-BSE calculations at $u=5,15$ for this phonon mode. At each of these atomic configurations, we obtain the excited state surface energy as the vertical excitation energy (from $GW$-BSE) plus the ground state energy (from DFT), in order
to obtain the total energy of the system in the excited
state. Solving the resulting system and after a few algebraic manipulations we find $u_{ES}\approx1$. 
While this
is a rough estimate, it is clear that the potential energy
surface minimum is far from the very large values of $u$
which are required in order to find the highly delocalized
excitons that appear for values of $u$ that are equal to $30$ or greater (see main manuscript Fig.\,3b).

\clearpage
\section{Tables of averages and standard deviations of exciton energies and radii}
\label{eq:values}

\begin{table}[b]
\centering
  \setlength{\tabcolsep}{6pt} 
\begin{tabular}{ccc}
\hline
$r_{\text{exc}}(T)$ (\AA)& singlet  & triplet \\
\hline
$0$\,K & $4.85\pm0.30$ &  $1.23\pm0.13$ \\
$100$\,K &  $5.04\pm0.17$ &   $1.23\pm0.10$\\
$200$\,K &  $4.98\pm0.21$ &  $1.32\pm0.07$ \\
$300$\,K & $6.96\pm1.25$ &  $1.26\pm0.13$ \\
\hline
\end{tabular}
\caption{Vibrational averages of exciton radii within the harmonic approximation - $2\times1\times1$ supercell.}
\label{table:rexc_averages_HA}
\end{table}

\begin{table}[b]
\centering
  \setlength{\tabcolsep}{6pt} 
\begin{tabular}{ccc}
\hline
$r_{\text{exc}}(T)$ (\AA)& singlet  & triplet \\
\hline
$100$\,K &  $5.90\pm0.65$ & $1.48\pm0.15$  \\
$300$\,K & $5.25\pm0.35$ &  $1.57\pm0.08$ \\
\hline
\end{tabular}
\caption{Vibrational averages of exciton radii within the anharmonic distribution - $2\times1\times1$ supercell.}
\label{table:rexc_averages_anhar}
\end{table}

\begin{table}[b]
\centering
  \setlength{\tabcolsep}{6pt} 
\begin{tabular}{cccccc}
\hline
$\text{E}(\text{S}_1)(T)$ (eV)& $-X$  & $-X/2$ & $\Gamma$ & $X/2$ & $X$ \\
\hline
$0$\,K &  $1.771\pm0.047$ & $1.735\pm0.056$  & $1.688\pm0.064$ & $1.737\pm0.056$ & $1.771\pm0.047$ \\
$300$\,K & $1.730\pm0.030$ &  $1.704\pm0.043$ & $1.663\pm0.053$ & $1.706\pm0.045$ & $1.733\pm0.033$\\
\hline
\end{tabular}
\caption{Singlet exciton energies at various points in reciprocal space, including the effect of $\Gamma$ phonons at $0$\,K and $300$\,K within the harmonic approximation.}
\label{table:E_singlet_HA}
\end{table}

\begin{table}[b]
\centering
  \setlength{\tabcolsep}{6pt} 
\begin{tabular}{cccccc}
\hline
$\text{E}(\text{T}_1)(T)$ (eV)& $-X$  & $-X/2$ & $\Gamma$ & $X/2$ & $X$ \\
\hline
$0$\,K &  $0.918\pm0.030$ & $0.909\pm0.027$  & $0.898\pm0.027$ & $0.910\pm0.028$ & $0.918\pm0.030$ \\
$300$\,K & $0.918\pm0.051$ &  $0.910\pm0.048$ & $0.899\pm0.045$ & $0.910\pm0.048$ & $0.918\pm0.052$\\
\hline
\end{tabular}
\caption{Triplet exciton energies at various points in reciprocal space, including the effect of $\Gamma$ phonons at $0$\,K and $300$\,K within the harmonic approximation.}
\label{table:E_triplet_HA}
\end{table}

\begin{table}[b]
\centering
  \setlength{\tabcolsep}{6pt} 
\begin{tabular}{ccc}
\hline
$\text{E}(\text{S}_1)(T)$ (eV)&  $\Gamma$ &  $X$ \\
\hline
$0$\,K &   $1.593\pm0.064$ &  $1.654\pm0.041$\\
$300$\,K &  $1.636\pm0.064$ & $1.712\pm0.061$\\
\hline
\end{tabular}
\caption{Singlet exciton energies at various points in reciprocal space, including the effects of $\Gamma$,$X$ phonons at $0$\,K and $300$\,K within the harmonic approximation.}
\label{table:E_singlet_har_sc}
\end{table}

\begin{table}[b]
\centering
  \setlength{\tabcolsep}{6pt} 
\begin{tabular}{cccc}
\hline
$\text{E}(\text{S}_1)(T)$ (eV)& $\Gamma$ & $X/2$ & $X$ \\
\hline
$100$\,K &  $1.713\pm0.037$ & $1.739\pm0.038$ & $1.780\pm0.043$ \\
$300$\,K &  $1.695\pm0.041$ & $1.719\pm0.039$ & $1.754\pm0.038$\\
\hline
\end{tabular}
\caption{Singlet exciton energies at various points in reciprocal space, including anharmonic effects and $\Gamma$ phonons at $100$\,K and $300$\,K.}
\label{table:E_singlet_anhar}
\end{table}

\begin{table}[b]
\centering
  \setlength{\tabcolsep}{6pt} 
\begin{tabular}{cccc}
\hline
$\text{E}(\text{T}_1)(T)$ (eV)&  $\Gamma$ & $X/2$ & $X$ \\
\hline
$100$\,K &   $0.885\pm0.078$ & $0.896\pm0.081$ & $0.904\pm0.083$\\
$300$\,K &  $0.872\pm0.063$ & $0.878\pm0.068$ & $0.891\pm0.068$\\
\hline
\end{tabular}
\caption{Triplet exciton energies at various points in reciprocal space, including anharmonic effects and $\Gamma$ phonons at $100$\,K and $300$\,K.}
\label{table:E_triplet_anhar}
\end{table}

\begin{table}[b]
\centering
  \setlength{\tabcolsep}{6pt} 
\begin{tabular}{cccc}
\hline
$\text{E}(\text{S}_1)(T)$ (eV)&  $\Gamma$ & $X/2$ & $X$ \\
\hline
$100$\,K &   $1.659\pm0.036$ & $1.692\pm0.033$ & $1.718\pm0.026$\\
$300$\,K &  $1.656\pm0.046$ & $1.675\pm0.044$ & $1.698\pm0.040$\\
\hline
\end{tabular}
\caption{Singlet exciton energies at various points in reciprocal space, including anharmonic effects and $\Gamma$,$X$ phonons at $100$\,K and $300$\,K.}
\label{table:E_singlet_anhar_sc}
\end{table}

\begin{table}[b]
\centering
  \setlength{\tabcolsep}{6pt} 
\begin{tabular}{cccc}
\hline
$\text{E}(\text{T}_1)(T)$ (eV)&  $\Gamma$ & $X/2$ & $X$ \\
\hline
$100$\,K &   $0.859\pm0.048$ & $0.870\pm0.037$ & $0.877\pm0.030$\\
$300$\,K &  $0.859\pm0.037$ & $0.865\pm0.044$ & $0.878\pm0.070$\\
\hline
\end{tabular}
\caption{Triplet exciton energies at various points in reciprocal space, including anharmonic effects and $\Gamma$,$X$ phonons at $100$\,K and $300$\,K.}
\label{table:E_triplet_anhar_sc}
\end{table}

\clearpage
\textbf{}
\bibliography{references}